\documentclass[9pt,twocolumn,twoside,dvipsnames]{opticajnl}

\journal{optica} 

\setboolean{shortarticle}{true}
\usepackage{amsmath, bm}
\usepackage[normalem]{ulem}

\title{Amplitude/Phase Retrieval for Terahertz Holography with Supervised and Unsupervised Physics-Informed Deep Learning}

\author[1,2,3,4]{Mingjun Xiang}
\author[3,*]{Hui Yuan}
\author[1,2]{Lingxiao Wang}
\author[1,2,*]{Kai Zhou}
\author[3]{Hartmut G. Roskos}

\affil[1]{Frankfurt Institute for Advanced Studies (FIAS), 60438 Frankfurt am Main, Germany}
\affil[2]{Xidian-FIAS International Joint Research Center, 60438 Frankfurt am Main, Germany}
\affil[3]{Physikalisches Institut, Goethe-Universität Frankfurt am Main, 60438 Frankfurt am Main, Germany}
\affil[4]{Xidian University, 710071 Xi'an, China}

\affil[*]{Corresponding author: yuan@physik.uni-frankfurt.de, zhou@fias.uni-frankfurt.de}
\begin{abstract}
Recently, digital holographic imaging techniques (including methods with heterodyne detection) have found increased attention in the terahertz (THz) frequency range. However, holographic techniques rely on the use of a reference beam in order to obtain phase information.
This letter investigates the potential of reference-free THz holographic imaging and proposes novel supervised and unsupervised deep learning (DL) methods for amplitude and phase recovery. The calculations incorporate Fresnel diffraction 
as prior knowledge. We first show that our \textit{unsupervised} dual network can predict amplitude and phase simultaneously, 
thus overcoming the limitation of previous studies which could only predict phase objects. This is demonstrated with synthetic 2D image data as well as with measured 2D THz diffraction images.
The advantage of unsupervised DL is that it can be used directly without labeling by human experts. We then address 
\textit{supervised} DL -- a concept of general applicability. We introduce training
with a database set of 2D images taken in the visible spectra range and modified by us numerically to emulate THz images. With this approach, we avoid the prohibitively time-consuming collection of a large number of THz-frequency images. The results obtained with both approaches represent the first steps towards fast holographic THz imaging with reference-beam-free low-cost power detection.
\end{abstract}

\setboolean{displaycopyright}{true}
\begin{document}

\maketitle





Imaging at THz frequencies (0.1–10~THz, wavelengths of 3~mm–30~$\mu$m) has received considerable attention in recent years. 
Application areas for THz imaging explored at present include non-destructive testing \cite{amenabar2013introductory}, quality monitoring \cite{ellrich2020terahertz}, security screening \cite{Federici_2005, Friederich2011}, biomedical imaging \cite{yang2016biomedical}, and sensing for robotics and vehicle control \cite{jasteh2015low}. The applied imaging modalities increasingly include coherent holographic approaches \cite{Valu21}, variously because they allow to capture of comparatively large scenes with good spatial resolution, offer the possibility of 3D scene reconstruction, may have a relaxed demand on the optical imaging optics, or lend themselves for advanced numerical image processing, e.g., exploiting sparsity effects. However, at lower THz frequencies, and there especially in the sub-1-THz frequency range, which is most relevant for many applications, one is  
forced to perform coherent detection as a serial rather than a parallel process \cite{nahata1996coherent, Yuan2019, Yuan2022}. The main reason is that power availability at these THz frequencies is a critical issue \cite{verghese1997optical}. While detector arrays (with pixel numbers limited to hundreds or at most a few thousand \cite{Valu21}) in principle are available, the limited power makes it difficult to provide a reference beam for multi-pixel interferometric or heterodyne phase measurements. The need for serial data recording renders
coherent imaging time-consuming and it is detrimental for image reconstruction, since phase distortions and noise introduced during signal recording affect the data quality strongly, especially in the case of weak signals \cite{wan2020terahertz,Yuan:19}. A powerful way to substantially reduce the power requirements, to drastically simplify the measurement system, and to accelerate the data acquisition process would open up, if 
the rebuilding of the phase information could be done computationally from the amplitude or intensity diffraction pattern obtained with the object-beam alone. Then it would become feasible to revert to array-based, reference-beam-free power detection for THz holography.

Conventional amplitude-phase recovery algorithms rely on iterative processing of diffraction patterns (DPs) recorded at different distances to improve the convergence and reliability of the reconstruction \cite{1972A,fienup1982phase,jaganathan2016phase}. The cost is the acquisition of large amounts of data and the manual searching for the convergence condition. Recently, rapidly evolving 
DL techniques offer a novel and efficient way to reconstruct images precisely \cite{goy2018low,zhang2018fast,hand2018phase}, including supervised DL methods which require plenty of labeled experimental data \cite{boominathan2018phase,deng2020learning,ju2018feature}, and unsupervised DL methods that are suitable for unlabeled data 
(such as PhysenNet \cite{2020Phase}). However, the expensive acquisition of THz images hinders naive supervised training with experimental data. 

This letter proposes two novel physics-informed DL methods -- one supervised and the other unsupervised -- for phase retrieval in THz holographic imaging, using only visible-light MNIST datasets \cite{MNIST} for training, and incorporating Fresnel diffraction as prior physical knowledge. 
Physics-informed DL provides a way to integrate 
mathematical physics models and data into the learning algorithms synergistically,ch that the calculations respect the physics laws symmetries of the modeled system 
\cite{raissi2019physics,karniadakis2021physics} and has been proven powerful in tackling inverse problems in physics  \cite{Pang:2016vdc,Shi:2021qri,Wang:2021jou}. The proposed supervised DL method requires longer training, which is however a one-off effort, leads to a high prediction speed after training, and makes the predictions fairly robust against noise in experimentally generated images (see later in Fig.~\ref{fig:false-color4}(f)).
The advantage of the unsupervised method is that it can be used directly both without labeling by human experts and 
training, 
and that it achieves simultaneous reconstruction of amplitude and phase, thus largely expanding the working range of previous unsupervised DL methods. 
\begin{figure}[ht]
\centering
\includegraphics[trim= 20 50 30 50, clip, width=\linewidth]{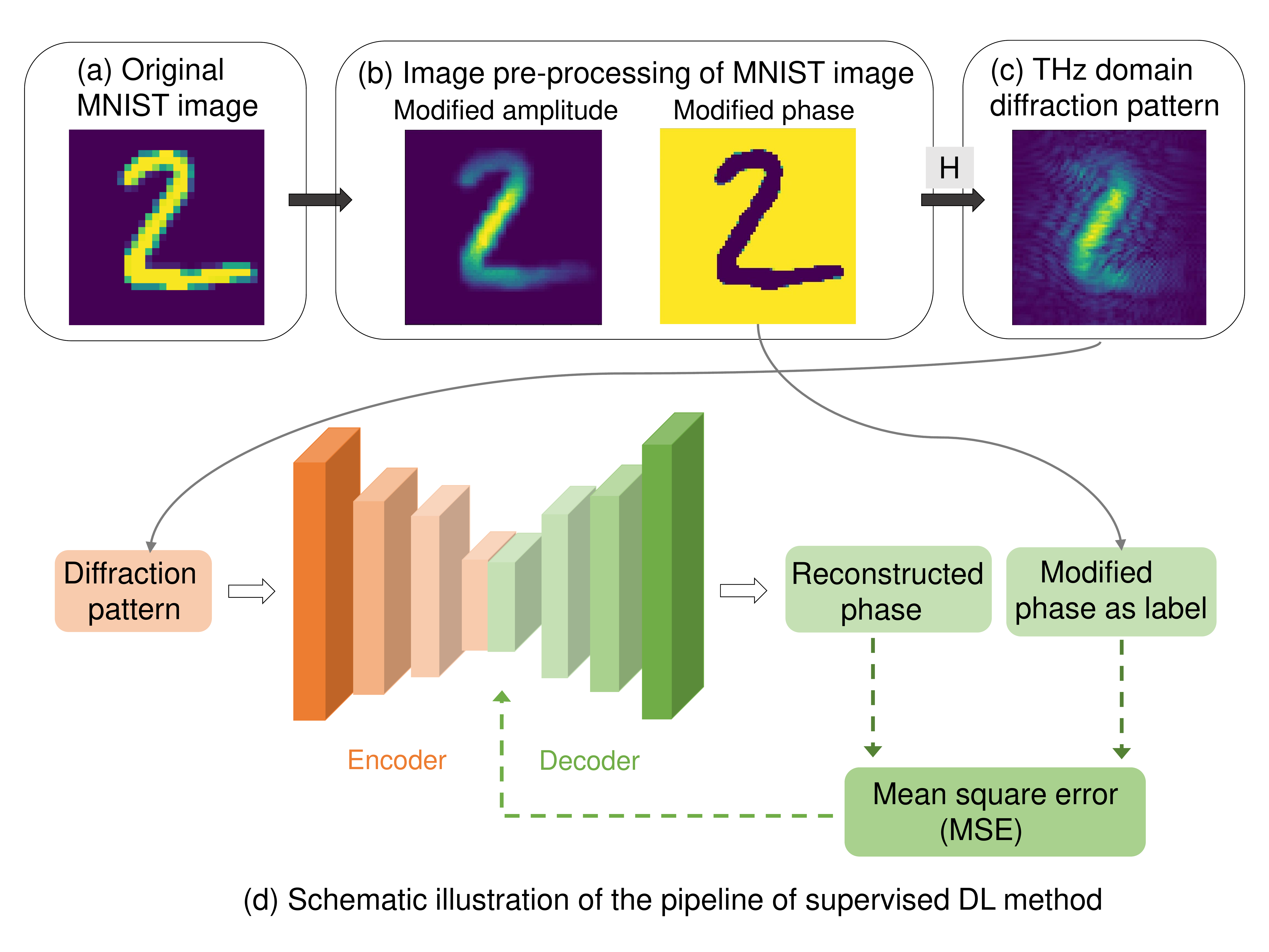}
\caption{Illustration of the supervised phase reconstruction model. (a) Random original image from the MNIST data set, (b) modified amplitude and phase image, (c) amplitude diffraction pattern in the THz domain calculated by the physical model. (d) Scheme of the convolutional neural network (CNN) of the supervised DL method. A measured diffraction pattern is the input of the CNN, and the output is a 
phase image. The mean squared error (MSE) between this phase reconstruction 
and the ground truth phase (i.e., the modified phase image in (b)) is taken as the loss function to optimize the CNN (see Supplementary for further details).}
\label{fig:false-color1}
\end{figure}

Figure~\ref{fig:false-color1} displays the flow chart of 
the proposed supervised phase reconstruction, illustrated for the use with handwritten digits. Lacking a sufficiently large data set recorded with THz radiation, 
we start from the public MNIST data set which was generated from thousands of photographs taken in the visible spectral range. 
The images from MNIST are then preprocessed to transpose them into effective THz images, as shown in Fig.~\ref{fig:false-color1}(a)-(c). 
The pre-processing includes (i) superimposing the Gaussian field-amplitude profile of a measured THz beam (recorded in the experimental setup discussed below) onto the images to obtain the modified amplitude, and (ii) determining the phase contour according to the object’s material and thickness to obtain the modified phase. 
We then employ the physical model $H$ (see below) to calculate the field-amplitude diffraction pattern of the effective THz image according to the Huygens–Fresnel principle \cite{goodman2005introduction}. This pattern is fed as input into the convolutional neural network (CNN), as shown in Fig.~1(d). Its architecture is inspired by U-Net \cite{alzubaidi2021review} and consists of a down-sampling path as well as a symmetric up-sampling path (see Supplementary for a detailed description). The output of the calculations are reconstructed phase images. In a subsequent step, the same network is applied for amplitude retrieval from the diffraction patterns.

The physical model, $H$, simulates the experimental THz imaging process. If a planar object is illuminated by a beam, the complex-valued field amplitude immediately behind the object can be written as
\begin{equation}
E_0(x,y,z=0) = A_0(x,y,0) e^{i\phi_0(x,y;0)},
\label{eq:refname1}
\end{equation}
where $A_0$ and $\phi_0$ are the amplitude and the phase at the object plane. Over a distance $d$, diffraction 
reshapes the field to \cite{goodman2005introduction}
\begin{equation}
E_d(x,y,z=d) = \iint \hat{E}_0(f_x,f_y)\,G\,e^{i2\pi(f_x x+f_y y)} df_x\,df_y,
\label{eq:refname2}
\end{equation}
where $G=e^{ikd\sqrt{1-\lambda^2f_x^2-\lambda^2f_y^2}}$ is the wave propagation 
function, $\lambda$ the wavelength, $\hat{E}_0$ the spatial Fourier transform of $E_0$ with $f_x=x/\lambda d$ and $f_y=y/\lambda d$ as the spatial frequencies in the $x$ and $y$ directions. 
The diffraction pattern 
is the field's absolute value 
\begin{equation}
A(x,y,z=d) = \left|E_d(x,y,z=d)\right| = H(\phi_0, A_0),
\label{eq:refname3}
\end{equation}
where $H(\cdot)$ represents the mapping function relating the object to the diffraction pattern $A$. The 
challenge is now to achieve 
an inverse mapping, ${H^{-1}(\cdot)}$, such that
\begin{equation}
\phi_0(x,y,0) = H^{-1}(A(x,y,z=d)).
\label{eq:refname4}
\end{equation}
The supervised method of Fig.~\ref{fig:false-color1} utilizes a parameterized network function $R_\theta$ ($\theta$ denoting the network weights and bias parameters) to approach the desired inverse mapping ${H^{-1}(\cdot)}$ via learning based upon
the labeled training set $S_T = {(A_k, \phi_k), k=1,2,...,K}$, thus solving the optimization problem of
\begin{equation}
R_{\theta^*} = \mathop{\arg\min}_{\theta} \|R_\theta(A_k)-\phi_k\|^2 \qquad\forall(\phi_k,A_k)\in S_T.
\label{eq:refname5}
\end{equation}
A corresponding inverse mapping reconstructs the amplitude information of the object, $A_0$. 

\begin{figure}[ht]
\centering
\includegraphics[trim= 40 100 30 70, clip, width=\linewidth]{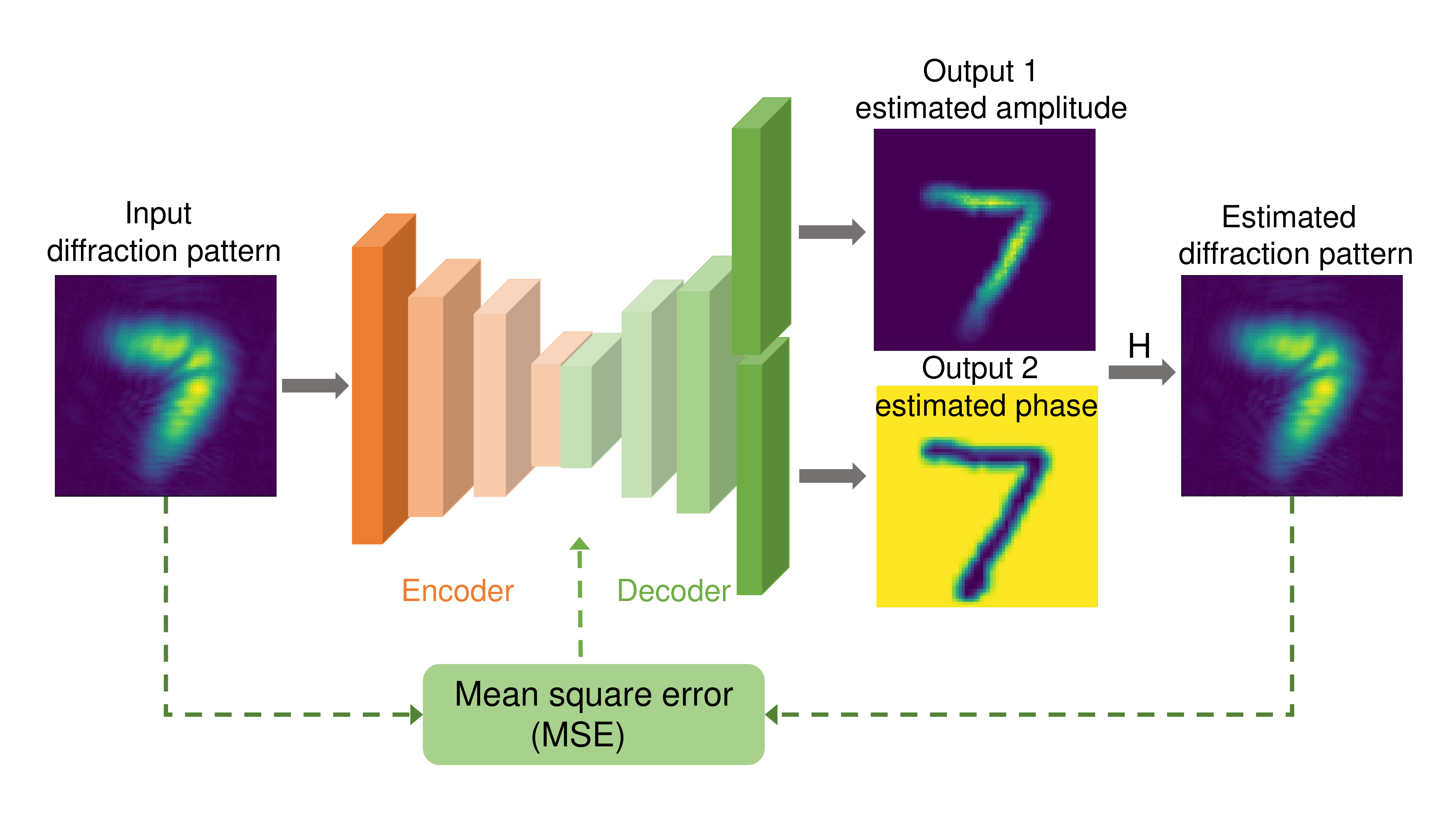}
\caption{Schematic illustration of the pipeline of unsupervised DL. The input to the NN is a measured field-amplitude diffraction pattern, the outputs are estimated phase and amplitude maps, which are then numerically propagated via the model $H$ to simulate the diffraction and measurement processes, yielding an estimated diffraction pattern. The MSE between the true and estimated patterns provides the loss function used to optimize the NN parameters (see Supplementary).}
\label{fig:false-color2}
\end{figure}
 
A different strategy is applied with the proposed unsupervised DL method shown in Fig.~\ref{fig:false-color2}.
Without the need for pre-training on large labeled datasets, 
it works directly on the amplitude $A(x, y, z = d)$, i.e., the propagated diffraction pattern of the object. 
Unlike PhysenNet, which was developed for visible-range images and only phase objects \cite{2020Phase}, the model proposed here is suitable for all holographic systems and has no object restrictions. The field-amplitude diffraction pattern is the only input to the 
CNN, which is designed to generate estimated phase and amplitude maps simultaneously (see Supplementary for a detailed description). The two output paths share the CNN's front layers in order to ensure that the same object information is analyzed, the path splits in two only at the backend. 
The model $H$ is then applied 
to convert the network outputs to an estimated diffraction pattern.
The MSE between the input and the estimated diffraction pattern is fed back to the network to optimize the weights and bias values via gradient descent. 
Thus, 
the retrieval of the phase is formulated as
\begin{equation}
R_{\theta^*} = \mathop{\arg\min}_{\theta} \|H(R_\theta(A))-A)\|^2 \,.
\label{eq:refname1}
\end{equation}
This will force the calculated diffraction pattern to converge to the measured pattern  $I$, as the iterative process proceeds. 

The pre-training and blind tests were performed on a PC with a sixteen-core 3.50-GHz CPU and 64 GB of RAM, using an Nvidia GeForce RTX 3080 GPU. On average, the pre-training of the supervised CNN took $\sim$4 h for 60,000 pairs of training data and 10,000 pairs of validation data during $\sim$100 training epochs. 
The inference time of the trained network for a hologram of $72 \times 72$ pixels was $< 0.1$~s. The training (optimization process) of the unsupervised NN (with phase and amplitude channel) took $\sim$1~min during $\sim$500 epochs. 

\begin{figure}[ht]
\centering
\includegraphics[width=\linewidth]{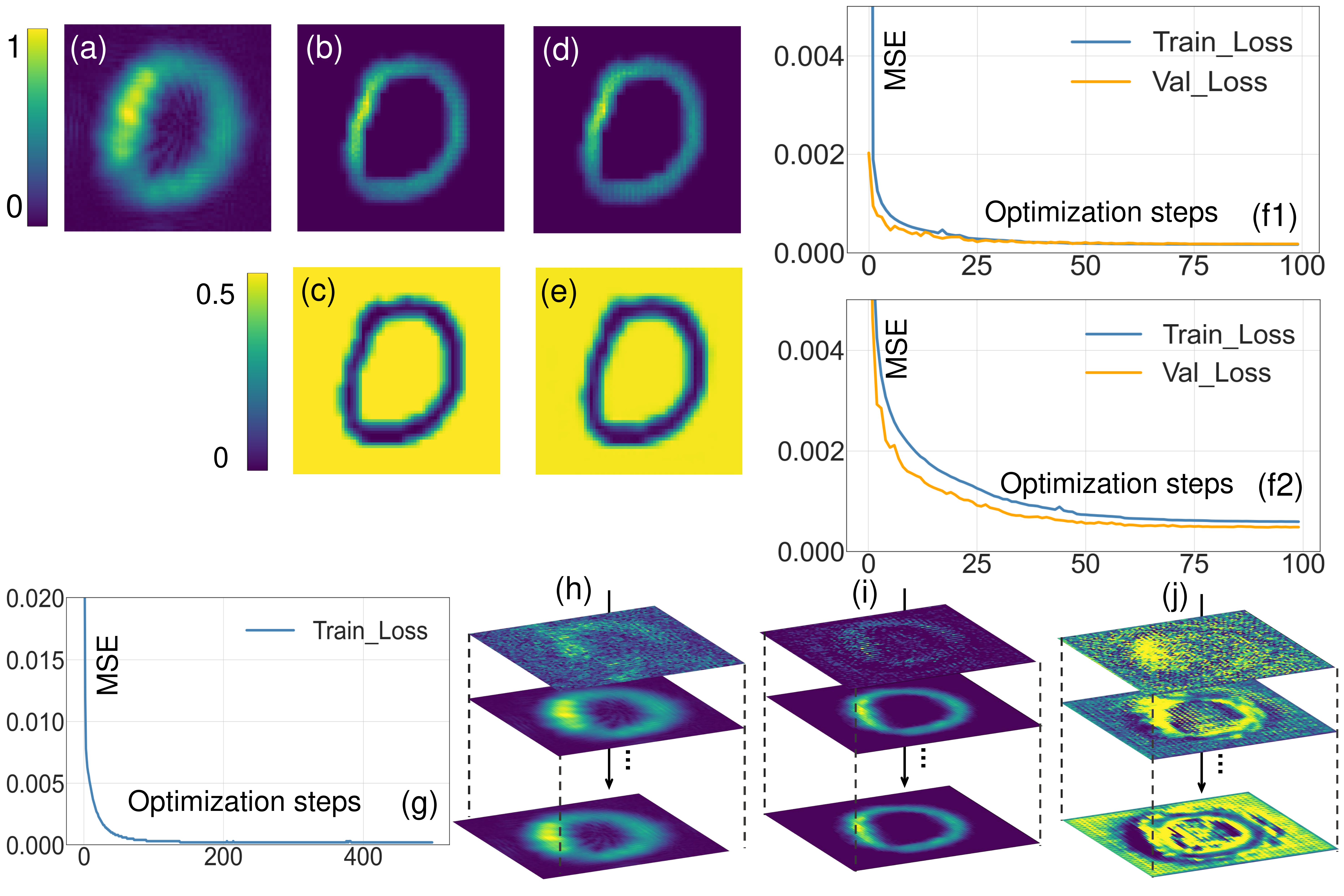}
\caption{Results obtained with the testing data set for supervised and unsupervised DL. (a) Diffraction pattern of a digit "0" from the testing data set. (b), (c) Ground truth of the object's amplitude and phase. (d)-(f) Supervised DL: Evolution of the MSE for amplitude (f1) and phase (f2) during pre-training (blue lines) with 60,000 images and testing with 10,000 ones (orange lines), and 
resulting predictions for the test "0" (d), (e). (g)-(j) Unsupervised DL: Optimization (training) curve (g) and evolution of the estimated diffraction pattern (h), amplitude (i), and phase (j) during optimization.}
\label{fig:false-color3}
\end{figure}

\begin{figure}[ht]
\centering
\includegraphics[width=\linewidth]{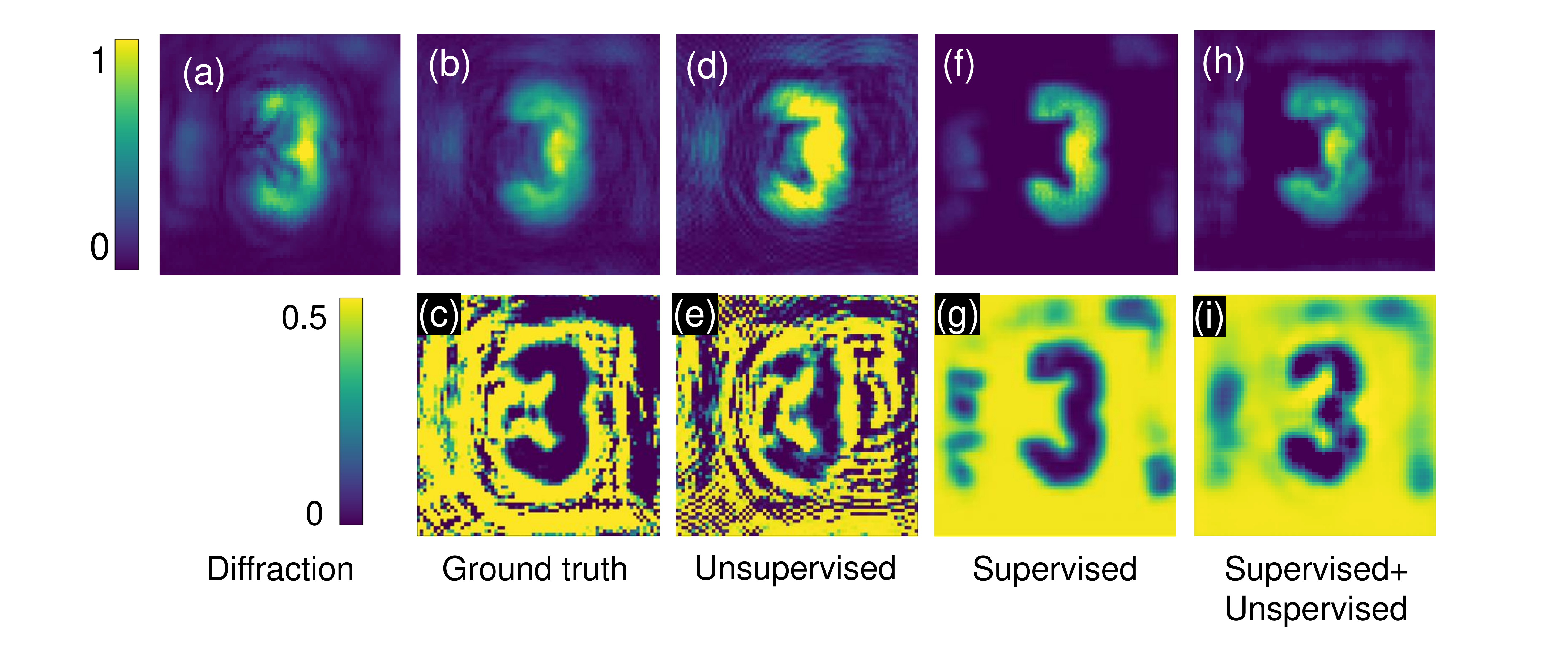}
\caption{Comparison of experimental and reconstruction results. 
(a) Amplitude diffraction pattern measured at a distance of 70~mm. (b), (c) Ground-truth amplitude and phase. (d)-(g) Reconstructed amplitude and phase maps, obtained by unsupervised DL ((d), (e)), respectively supervised DL ((f), (g)). (h), (i) Amplitude and phase images reconstructed by the sequential combination of the two methods.}
\label{fig:false-color4}
\end{figure}

We demonstrate the performance of the proposed supervised and unsupervised methods with data derived from simulation (THz-emulated MNIST) and experiment. 
Fig.~\ref{fig:false-color3} shows results obtained with the two methods for a randomly chosen MNIST number (``0'').
Note that the diffraction pattern as shown in Fig.~\ref{fig:false-color3}(a) is the only input to both methods. In the case of the supervised approach, the networks are first pre-trained to represent the mapping function between the amplitude 
diffraction pattern and the objects' phases and amplitudes. The corresponding learning curves in Fig.~\ref{fig:false-color3}(f1) and f(2) show convergence to low validation loss values after $\sim$50 epochs,
indicating that the networks have not become overfitted by the training data set. 
We obtain a validation loss (MSE) of 0.00018 and 0.00048 after 100 epochs 
for amplitude and phase, respectively. 
The reconstructed amplitude and phase of the test object are shown in Fig.~\ref{fig:false-color3}(d) and (e), with the ground truth for comparison in (b) and (c). The MSE between the ground truth and the reconstruction is 0.00016 and 0.00011 for phase and amplitude, respectively. 

As stated above, the proposed unsupervised DL method works directly on 
diffraction data without prior training. The optimization process can be monitored by the 
loss curve (MSE between the input diffraction pattern and the one 
calculated from the CNN-estimated phase and amplitude data) which is displayed in Fig.~\ref{fig:false-color3}(g). As the optimization goes on, the MSE drops to $10^{-6}$ after 500 epochs, yielding 
the concurrently updated diffraction, amplitude and phase patterns shown in Fig.~\ref{fig:false-color3}(h)-(j). 
It is found that the MSE between the ground-truth phase and the predicted phase is 0.06, while the MSE between the ground-truth amplitude and predicted amplitude is 0.00029. Compared to the supervised DL method, the phase prediction performance is slightly worse, while the amplitude prediction is comparably good. This trend is confirmed with other test objects. In more ambiguous cases, supervised DL is found to be clearly superior to unsupervised DL (see Supplementary).
Table~\ref{tab:comp} summarizes the findings.
\begin{table}[htbp]
\centering
\caption{\bf Comparison of the two methods}
\begin{tabular}{cccc}
\hline
method & accuracy & reliability & pre-training \\
\hline
supervised DL & higher & higher & yes\\
unsupervised DL & lower & lower & no \\
\hline
\end{tabular}
  \label{tab:comp}
\end{table}

To validate the proposed methods by experiments, we performed 
coherent heterodyne THz measurements at 300~GHz with two frequency-locked electrical multiplier chains (see also Supplementary). The objects consisted of thin metal screens with cut-outs in the form of digits. 
The collimated THz beam 
illuminated the object, the transmitted diffracted wave was detected coherently 70~mm behind the object by a raster-scanned single-pixel TeraFET detector \cite{Ikamas2018}. 
The detector and the second emitter, which provided the local-oscillator signal, were co-located on a 2D translation stage. 
From the measured complex-valued diffraction pattern ($E_d (z=d)$), the object's amplitude ($A_0$) and phase ($\phi_0$) were reconstructed by back-propagation via the inverse operation of Eq.~\ref{eq:refname2}. $A_0$, $\phi_0$ represent the ground truth for the DL evaluation of the image represented by $A(z=d)$. 



The measured $A(z=d)$ diffraction pattern and the ground truth together with predictions from the DL methods are shown in Fig.~\ref{fig:false-color4}. 
We reconstructed the amplitude and phase in three different ways, by unsupervised and supervised DL (Fig.~\ref{fig:false-color4}(d)-(g)) and 
by a combination of both in a sequential process where the prediction obtained by supervised DL served as input to the following unsupervised process, yielding the final reconstruction shown in Fig.~\ref{fig:false-color4}(h), (i).
The MSE values 
(referenced to the ground truth) for the unsupervised method are 0.006 (0.05) for the reconstructed amplitude (phase). The corresponding values for supervised DL are 0.0084 (0.0249). The combined method yields the best MSE values: 0.0035 (0.0230) for amplitude (phase). The image exhibits sharper edges and better contrast, and is closer to the ground truth. We attribute the reason for this considerable improvement to a suppression of the influence of noise, which is present in the measured images, by the supervised DL step. This denoising effect invites further studies in the future. 


In conclusion, we have proposed supervised and unsupervised deep-learning methods, and a combination thereof, for the reconstruction of 2D terahertz images, starting from amplitude diffraction patterns and deriving the amplitude and phase in the object plane.
As it is next to impossible to generate by coherent measurements a THz image database of sufficient size for pre-training of the neural networks, 
a large set of publicly accessible visible-range images was converted numerically into emulated complex-valued THz images. 
The neural network trained with these images achieves fast and reliable image reconstruction with both emulated and experimentally measured THz amplitude images. 
Regarding the proposed unsupervised method, a dual output layer, integrating a physical model of the wave propagation, has been introduced to simultaneously predict the amplitude and phase information. The best reconstruction performance with measured image data is achieved by the sequential application of both methods: The first step with the supervised method eliminates much of the noise of the raw images and provides an already good image estimate for the subsequent unsupervised method. At least for sufficiently simple scenes, our findings confirm the potential of phase retrieval from amplitude images of THz diffraction patterns.



\begin{backmatter}
\bmsection{Funding} The work was supported by XF-IJRC (M. Xiang), the German Research Foundation DFG with grant RO 770/48-1 (H. Yuan), the BMBF under ErUM-Data (K. Zhou), the AI grant of SAMSON AG, Frankfurt (K. Zhou, L. Wang).
\bmsection{Disclosures} The authors declare no conflicts of interest.
\bmsection{Data Availability Statement} Raw data will be made available upon reasonable request.
\bmsection{Supplemental document} Supporting content provided. 
\end{backmatter}

\bibliography{ref}

\begin{thebibliography}{10}
\newcommand{\enquote}[1]{``#1''}

\bibitem{amenabar2013introductory}
I.~Amenabar, F.~Lopez, and A.~Mendikute, \enquote{In introductory review to thz
  non-destructive testing of composite mater,} {\protect\JournalTitle{Journal
  of Infrared, Millimeter, and Terahertz Waves}} \textbf{34}, 152--169 (2013).

\bibitem{ellrich2020terahertz}
F.~Ellrich, M.~Bauer, N.~Schreiner, A.~Keil, T.~Pfeiffer, J.~Klier, S.~Weber,
  J.~Jonuscheit, F.~Friederich, and D.~Molter, \enquote{Terahertz quality
  inspection for automotive and aviation industries,}
  {\protect\JournalTitle{Journal of Infrared, Millimeter, and Terahertz Waves}}
  \textbf{41}, 470--489 (2020).

\bibitem{Federici_2005}
J.~F. Federici, B.~Schulkin, F.~Huang, D.~Gary, R.~Barat, F.~Oliveira, and
  D.~Zimdars, \enquote{{THz} imaging and sensing for security
  applications{\textemdash}explosives, weapons and drugs,}
  {\protect\JournalTitle{Semiconductor Science and Technology}} \textbf{20},
  S266--S280 (2005).

\bibitem{Friederich2011}
F.~Friederich, W.~von Spiegel, M.~Bauer, F.~Z. Meng, M.~D. Thomson, S.~Boppel,
  A.~Lisauskas, B.~Hils, V.~Krozer, A.~Keil, T.~Löffler, R.~Henneberger, A.~K.
  Huhn, G.~Spickermann, P.~H. Bolívar, and H.~G. Roskos, \enquote{Thz active
  imaging systems with real-time capabilities,} {\protect\JournalTitle{IEEE
  Transactions on Terahertz Science and Technology}} \textbf{1}, 183--200
  (2011).

\bibitem{yang2016biomedical}
X.~Yang, X.~Zhao, K.~Yang, Y.~Liu, Y.~Liu, W.~Fu, and Y.~Luo,
  \enquote{Biomedical applications of terahertz spectroscopy and imaging,}
  {\protect\JournalTitle{Trends in Biotechnology}} \textbf{34}, 810--824
  (2016).

\bibitem{jasteh2015low}
D.~Jasteh, M.~Gashinova, E.~Hoare, T.-Y. Tran, N.~Clarke, and M.~Cherniakov,
  \enquote{Low-thz imaging radar for outdoor applications,} in \emph{2015 16th
  International Radar Symposium (IRS),}  (IEEE, 2015), pp. 203--208.

\bibitem{Valu21}
G.~Valušis, A.~Lisauskas, H.~Yuan, W.~Knap, and H.~G. Roskos, \enquote{Roadmap
  of terahertz imaging 2021,} {\protect\JournalTitle{Sensors}} \textbf{21},
  4092 (2021).

\bibitem{nahata1996coherent}
A.~Nahata, D.~H. Auston, T.~F. Heinz, and C.~Wu, \enquote{Coherent detection of
  freely propagating terahertz radiation by electro-optic sampling,}
  {\protect\JournalTitle{Applied Physics Letters}} \textbf{68}, 150--152
  (1996).

\bibitem{Yuan2019}
H.~Yuan, D.~Voß, A.~Lisauskas, D.~Mundy, and H.~G. Roskos, \enquote{{3D}
  {Fourier} imaging based on {2D} heterodyne detection at {THz} frequencies,}
  {\protect\JournalTitle{APL Photonics}} \textbf{4}, 106108 (2019).

\bibitem{Yuan2022}
H.~Yuan, A.~Lisauskas, M.~D. Thomson, and H.~G. Roskos, \enquote{{600-GHz}
  {Fourier} imaging based on heterodyne detection at the 2nd sub-harmonic,}
  {\protect\JournalTitle{IEEE Transactions on Terahertz Science and
  Technology}} p. submitted (2022).

\bibitem{verghese1997optical}
S.~Verghese, K.~McIntosh, and E.~Brown, \enquote{Optical and terahertz power
  limits in the low-temperature-grown gaas photomixers,}
  {\protect\JournalTitle{Applied Physics Letters}} \textbf{71}, 2743--2745
  (1997).

\bibitem{wan2020terahertz}
M.~Wan, J.~J. Healy, and J.~T. Sheridan, \enquote{Terahertz phase imaging and
  biomedical applications,} {\protect\JournalTitle{Optics \& Laser Technology}}
  \textbf{122}, 105859 (2020).

\bibitem{Yuan:19}
H.~Yuan, M.~Wan, A.~Lisauskas, J.~T. Sheridan, and H.~G. Roskos,
  \enquote{300-{GHz} holography with heterodyne detection,} in \emph{Digital
  Holography and Three-Dimensional Imaging 2019,}  (Optica Publishing Group,
  2019), p. Th3A.21.

\bibitem{1972A}
R.~W. Gerchberg and W.~O. Saxton, \enquote{A practical algorithm for the
  determination of phase from image and diffraction plane pictures,}
  {\protect\JournalTitle{Optik}} \textbf{35}, 237--250 (1972).

\bibitem{fienup1982phase}
J.~R. Fienup, \enquote{Phase retrieval algorithms: a comparison,}
  {\protect\JournalTitle{Applied Optics}} \textbf{21}, 2758--2769 (1982).

\bibitem{jaganathan2016phase}
K.~Jaganathan, Y.~C. Eldar, and B.~Hassibi, \enquote{Phase retrieval: An
  overview of recent developments,} {\protect\JournalTitle{Optical Compressive
  Imaging}} pp. 279--312 (2016).

\bibitem{goy2018low}
A.~Goy, K.~Arthur, S.~Li, and G.~Barbastathis, \enquote{Low photon count phase
  retrieval using deep learning,} {\protect\JournalTitle{Physical Review
  Letters}} \textbf{121}, 243902 (2018).

\bibitem{zhang2018fast}
G.~Zhang, T.~Guan, Z.~Shen, X.~Wang, T.~Hu, D.~Wang, Y.~He, and N.~Xie,
  \enquote{Fast phase retrieval in off-axis digital holographic microscopy
  through deep learning,} {\protect\JournalTitle{Optics Express}} \textbf{26},
  19388--19405 (2018).

\bibitem{hand2018phase}
P.~Hand, O.~Leong, and V.~Voroninski, \enquote{Phase retrieval under a
  generative prior,} {\protect\JournalTitle{Advances in Neural Information
  Processing Systems}} \textbf{31} (2018).

\bibitem{boominathan2018phase}
L.~Boominathan, M.~Maniparambil, H.~Gupta, R.~Baburajan, and K.~Mitra,
  \enquote{Phase retrieval for {Fourier} ptychography under varying amount of
  measurements,} {\protect\JournalTitle{arXiv:1805.03593}}  (2018).

\bibitem{deng2020learning}
M.~Deng, S.~Li, A.~Goy, I.~Kang, and G.~Barbastathis, \enquote{Learning to
  synthesize: robust phase retrieval at low photon counts,}
  {\protect\JournalTitle{Light: Science \& Applications}} \textbf{9}, 1--16
  (2020).

\bibitem{ju2018feature}
G.~Ju, X.~Qi, H.~Ma, and C.~Yan, \enquote{Feature-based phase retrieval
  wavefront sensing approach using machine learning,}
  {\protect\JournalTitle{Optics Express}} \textbf{26}, 31767--31783 (2018).

\bibitem{2020Phase}
W.~Fei, Y.~Bian, H.~Wang, L.~Meng, and G.~Situ, \enquote{Phase imaging with an
  untrained neural network,} {\protect\JournalTitle{Light: Science \&
  Applications}} \textbf{9}, 77 (2020).

\bibitem{MNIST}
Modified National Institute of Standards and Technology database of handwritten
  2D digits, {http://yann.lecun.com/exdb/mnist/}.

\bibitem{raissi2019physics}
M.~Raissi, P.~Perdikaris, and G.~E. Karniadakis, \enquote{Physics-informed
  neural networks: A deep learning framework for solving forward and inverse
  problems involving nonlinear partial differential equations,}
  {\protect\JournalTitle{Journal of Computational Physics}} \textbf{378},
  686--707 (2019).

\bibitem{karniadakis2021physics}
G.~E. Karniadakis, I.~G. Kevrekidis, L.~Lu, P.~Perdikaris, S.~Wang, and
  L.~Yang, \enquote{Physics-informed machine learning,}
  {\protect\JournalTitle{Nature Reviews Physics}} \textbf{3}, 422--440 (2021).

\bibitem{Pang:2016vdc}
L.-G. Pang, K.~Zhou, N.~Su, H.~Petersen, H.~St\"ocker, and X.-N. Wang,
  \enquote{{An equation-of-state-meter of quantum chromodynamics transition
  from deep learning},} {\protect\JournalTitle{Nature Commun.}} \textbf{9}, 210
  (2018).

\bibitem{Shi:2021qri}
S.~Shi, K.~Zhou, J.~Zhao, S.~Mukherjee, and P.~Zhuang, \enquote{{Heavy quark
  potential in the quark-gluon plasma: Deep neural network meets lattice
  quantum chromodynamics},} {\protect\JournalTitle{Phys. Rev. D}} \textbf{105},
  014017 (2022).

\bibitem{Wang:2021jou}
L.~Wang, S.~Shi, and K.~Zhou, \enquote{{Reconstructing spectral functions via
  automatic differentiation},} {\protect\JournalTitle{Phys. Rev. D}}
  \textbf{106}, L051502 (2022).

\bibitem{goodman2005introduction}
J.~W. Goodman, \enquote{Introduction to fourier optics, roberts \& co,}
  {\protect\JournalTitle{Publishers, Englewood, Colorado}}  (2005).

\bibitem{alzubaidi2021review}
L.~Alzubaidi, J.~Zhang, A.~J. Humaidi, A.~Al-Dujaili, Y.~Duan, O.~Al-Shamma,
  J.~Santamar{\'\i}a, M.~A. Fadhel, M.~Al-Amidie, and L.~Farhan,
  \enquote{Review of deep learning: Concepts, cnn architectures, challenges,
  applications, future directions,} {\protect\JournalTitle{Journal of Big
  Data}} \textbf{8}, 1--74 (2021).

\bibitem{Ikamas2018}
K.~Ikamas, D.~Čibiraitė, A.~Lisauskas, M.~Bauer, V.~Krozer, and H.~G. Roskos,
  \enquote{Broadband terahertz power detectors based on 90-nm silicon {CMOS}
  transistors with flat responsivity up to 2.2~{THz},}
  {\protect\JournalTitle{IEEE Electron Device Letters}} \textbf{39}, 1413--1416
  (2018).

\end{thebibliography}

\bibliographyfullrefs{ref}

\end{document}